\documentclass{ws-procs9x6}
\newcommand{\beq}{\begin{equation}}
\newcommand{\eeq}{\end{equation}}
\newcommand{\bea}{\begin{eqnarray}}
\newcommand{\eea}{\end{eqnarray}}

\def\half{\frac{1}{2}}
\def\nubar{\overline{\nu}}
\def\Acal{\mathcal{A}}

\def\GMem{G_M}

\def\GENC{G_E^{NC}}
\def\GMNC{G_M^{NC}}
\def\GANC{G_A^{NC}}
\def\GACC{G_A}

\def\GAs{G_A^s}
\def\GMs{G_M^s}

\begin{document}

\title{Strange form factors in neutrino scattering}

\author{Wanda M. Alberico}

\address{Dipartimento di Fisica Teorica, Universit\`a  di Torino \\
and INFN, Sezione di Torino,\\ 
Via P. Giuria 1, I--10125 Torino, Italy\\ 
E-mail: alberico@to.infn.it}

\author{Chiara Maieron}
\address{Departamento de Fisica Atomica, Molecular y Nuclear,\\ 
Universidad de Sevilla, \\
Avda. Reina Mercedes s/n, E--41013 Sevilla, Spain \\
E-mail: chiara@nucle.us.es}  


\maketitle

\abstracts{The possibility to determine the
axial strange form factor of the nucleon from neutrino
scattering experiments is studied. The existing experimental 
information is reviewed and several related observables which could be 
measured in the near future at new neutrino facilities are studied
in detail.
Elastic scattering from $S=T=0$ nuclei is also briefly considered.}

\section{Introduction}
The determination of the strange quark contribution to
the nucleonic axial and vector weak currents has
raised large and continuous interest in theoretical and experimental
physics, especially after the measurements of the polarised 
structure function $g_1^p$ of the proton, which indicated a non--zero 
contribution of the strange quark to the proton spin.

In recent years, intermediate energy physics
has been mainly focused on parity violating electron scattering
and on the strange {\it vector} current, while the strange contribution
to the nucleon {\it axial} current has been widely investigated only at 
higher energies.

The measurement of the structure function
$g_1^p(x)$ in polarised deep inelastic scattering
can be used 
to determine the one nucleon 
matrix elements of the axial quark current~\cite{alb02}
\begin{equation}
\langle p,\,s | \overline{q}\gamma^{\alpha}\gamma^5 q |
p,\, s \rangle = 2 M s^{\alpha} g_A^q \;.
\label{eq:gau}
\end{equation}
This can be obtained by combining the QCD sum rule 
$\Gamma_1^p = \int_{0}^{1} dx g_1^p(x)$ , which in the naive
quark parton model has the following flavour structure
\begin{equation}
\Gamma_1^p = \frac{1}{2}\left( \frac{4}{9}g_A^u +
 \frac{1}{9}g_A^d + \frac{1}{9}g_A^s \right) \;,
\label{eq:gamma1p}
\end{equation}
with the relations
\begin{equation}
g_A = g_A^u - g_A^d\;,
\label{eq:ga}
\end{equation}
based on the isotopic $SU(2)$ invariance of strong interactions,
and 
\begin{equation}
3F - D =  g_A^u + g_A^d -2 g_A^s\;,
\label{eq:hyper}
\end{equation}
based on the $SU(3)_f$ symmetry.
Here the axial constant $g_A = 1.2573 \pm 0.0028$ is obtained
from neutron beta decay, while the constants $F$ and $D$ come from 
the measurements of semileptonic decays of hyperons.

Despite the continuous improvements in both
experimental accuracy and theoretical calculations, 
the determination of the values of $g_A^ {u,d,s}$  is 
still subject to several strong assumptions, 
such as the small $x$ extrapolation of  
$g_1^p(x)$, the QCD corrections used to relate its
first moment $\Gamma_1^p$
to the constants $g_A^q$ and the  $SU(3)_f$ invariance assumed in
eq. (\ref{eq:hyper}).

It is therefore interesting to look for alternative methods
for measuring $g_A^s$, which do not rely on these same assumptions.

It has long been recognised that neutrino scattering
\cite{kap88} can be a powerful
tool for this investigation, but up to now the poor precision of 
the experimental data has 
not allowed the extraction, from them, of unambiguous results. 

However, due to the large interest in neutrino physics
raised by the recent results on neutrino oscillations,
new neutrino facilities are being constructed, 
which could reach the required experimental 
accuracy for using  neutrino scattering processes as a precise
probe of $g_A^s$. 
 
In this contribution, after presenting the relevant formalism 
for the description of neutrino  scattering, we will
review the existing  information on the nucleon strange form 
factors obtained from these processes and we will explore the possibility
to obtain new, definite measurements of $g_A^s$ in the near future.
Since the main focus here is on the strangeness content of the nucleon,
we will mostly consider scattering processes on free nucleons;
a detailed analysis of nuclear structure effects can be found
elsewhere~\cite{alb02}.

A short discussion of elastic neutrino scattering on $S=T=0$ nuclei
will be also presented.

\section{Formalism}

Let us consider the neutral current (NC) processes
\begin{equation}
\nu_{\mu} (\overline{\nu}_{\mu}) + N
\rightarrow  \nu_{\mu} (\overline{\nu}_{\mu}) + N \;.
\label{eq:NC}
\end{equation}
In the Standard Model and considering the contributions of
$u$, $d$ and $s$ quarks only, the weak nuclear current involved in
these processes can be written in the form:
\begin{eqnarray}
\label{eq:jnc}
J_{\alpha}^{NC} &=& V_{\alpha}^{NC} + A_{\alpha}^{NC} 
\\
&=& V_{\alpha}^3 -2 \sin^2(\theta_W) J_{\alpha}^{em} -
\frac{1}{2}V_{\alpha}^s + A_{\alpha}^3 - \frac{1}{2}A_{\alpha}^s \;,    
\nonumber
\end{eqnarray}
where the isovector polar and axial vector currents are given by
\begin{eqnarray}
V_{\alpha}^3 &=& \frac{1}{2}\left\{\overline{U}\gamma_{\alpha}U -
 \overline{D}\gamma_{\alpha}D\right\}  
\nonumber 
\\
A_{\alpha}^3 &=& \frac{1}{2}\left\{
\overline{U}\gamma_{\alpha}\gamma_{5}U -
 \overline{D}\gamma_{\alpha}\gamma_{5}D\right\}  \;,  
\label{eq:isovector}
\end{eqnarray}
$J_{\alpha}^{em}$ is the electromagnetic current
and the strange currents $V_{\alpha}^s$ and $A_{\alpha}^s$ 
are defined as
\begin{eqnarray}
V_{\alpha}^s &=& \overline{S}\gamma_{\alpha}S 
\nonumber 
\\
A_{\alpha}^s &=& \overline{S}\gamma_{\alpha}\gamma_{5}S\;.
\label{eq:strange}
\end{eqnarray}

Complementary to the NC processes (\ref{eq:NC}) are 
the Charged Current (CC) reactions
\begin{eqnarray}
\nu_{\mu} + n
&\rightarrow&  \mu^{-} + p
\nonumber \\
\overline{\nu}_{\mu} + p
&\rightarrow&  \mu^{+} + n \;,
\label{eq:CC}
\end{eqnarray}
with the corresponding currents 
$J_{\alpha}^{CC} = V_{ud} 
\overline{U} \gamma_{\alpha}(1 + \gamma_5) D$ and 
$(J_{\alpha}^{CC})^\dagger$, $V_{ud}$ being the $ud$
Cabibbo--Kobayashi--Maskawa matrix element.

The cross sections for the processes (\ref{eq:NC}) depend on
the matrix elements of the weak neutral current (\ref{eq:jnc}),
taken between one nucleon states of initial and final
momentum $p$ and $p'$, respectively. 
Their nucleon structure content can be parameterised
in terms of three NC form factors, according to
\begin{eqnarray}
_{p(n)}\langle p^{\prime} | V_{\alpha}^{NC} | p \rangle _{p(n)}
&=&
\overline{u}(p^{\prime})
\left[ \gamma_{\alpha} F_1^{NC; p(n)}(Q^2) +
\frac{i}{2M} \sigma_{\alpha \beta}q^\beta F_2^{NC; p(n)}(Q^2)
\right] u(p)
\nonumber 
\\
_{p(n)}\langle p^{\prime} | A_{\alpha}^{NC} | p \rangle _{p(n)}
&=&
\overline{u}(p^{\prime})
\gamma_{\alpha}\gamma_5 G_A^{NC; p(n)}(Q^2) u(p)\;,
\label{eq:ffnc}
\end{eqnarray}
where $q$ is the four-momentum transfer and $Q^2 = -q_{\alpha}^2$.

Using eq. (\ref{eq:jnc}), the form factors $F_{1,2}^{NC;n(p)}$
can be written in the following form:
\begin{eqnarray}
F_{1,2}^{NC;p(n)}(Q^2) &=&  
\pm \frac{1}{2}
\left[F_{1,2}^p(Q^2) - F_{1,2}^n(Q^2) \right] -
\nonumber\\
&&-2\sin^2(\theta_W) F_{1,2}(Q^2) - \frac{1}{2}F_{1,2}^s(Q^2)\;,
\label{eq:ffnc2}
\end{eqnarray}
where $F_{1,2}^{p(n)}$ are the proton (neutron) Pauli and Dirac
electromagnetic form factors and the plus and minus sign refer to
the proton and neutron, respectively.
Equivalently, NC Sachs form factors can be used, whose expressions are,
correspondingly:
\begin{eqnarray}
G_{E,M}^{NC;p(n)}(Q^2) &=&  
\pm \frac{1}{2}
\left[G_{E,M}^p(Q^2) - G_{E,M}^n(Q^2) \right] -
\nonumber\\
&&-2\sin^2(\theta_W) G_{E,M}^{p(n)}(Q^2) - \frac{1}{2}G_{E,M}^s(Q^2)\;.
\label{eq:gemnc}
\end{eqnarray}
Moreover, using the isotopic invariance of strong interactions,
the one nucleon NC axial matrix elements can be written
as
\begin{equation}
G_A^{NC;p(n)}(Q^2) =  
\pm \frac{1}{2} G_A(Q^2) -  \frac{1}{2} G_A^s(Q^2)\;, 
\label{eq:ganc}
\end{equation}
where, again, the plus (minus) sign refers to the proton (neutron),
and $ G_A(Q^2)$ is the { usual} CC axial form factor,
measured in the processes (\ref{eq:CC}).

From eqs. (\ref{eq:ffnc2})--(\ref{eq:ganc}) we can see that
the NC nucleon current can be expressed in terms of known
(electromagnetic and CC) form factors, plus unknown contributions
coming from strange quarks.
In particular the neutrino--nucleon cross sections 
turn out to be very
sensitive to the NC axial form factor and thus, in combination
with the measurement of the  strange vector form factors $F_{1,2}^s$ (or
$G_{E,M}^s$) from parity violating electron scattering, 
their measurement can be used to extract $G_A^s$.

We remark that, for the same reason, an accurate knowledge
of the CC axial form factor and of its $Q^2$ dependence is essential
to be able to separate $G_A^s$. Thus, when planning future 
neutrino experiments it would be important to consider the possibility
of performing high precision measurements of CC cross sections
as well as NC ones. 

Since very little is known about the $Q^2$ dependence of the strange form 
factors, some assumptions must be done when studying their effects
in the cross sections. In the following we will assume, as it is generally 
done, that the strange form factors have the same $Q^2$ behaviour of the
corresponding non--strange ones. In particular a dipole form will
be assumed for $G_M^s$, with cutoff mass $M_V = 0.84$ GeV and 
$G_M^s(0) = \mu_s$, while for $G_A^s$ the same cutoff mass as for the 
CC axial form factor $G_A$ is used, with $G_A^s(0) = g_A^s$.

In terms of NC form factors, the differential cross section
for the processes (\ref{eq:NC}) has the following explicit form:
\bea
&&\left(\frac{d\sigma}{dQ^2}\right)^{NC}_{\nu(\nubar)} = 
\nonumber\\
&& =
\frac{G_F^2}{2\pi}\left[\half y^2(\GMNC)^2 +\left(1-y-\frac{M}{2E}y\right)
\frac{\displaystyle{(\GENC)^2+\frac{E}{2M}y(\GMNC)^2}}
{\displaystyle{1+\frac{E}{2M}y}}\right.
\nonumber\\
&&\left.
+\left(\half y^2+1-y+\frac{M}{2E}y\right)(\GANC)^2
\pm 2y\left(1-\half y\right)\GMNC\GANC\right]\,.
\label{eq:nc_cross}
\eea
Here
$y=\displaystyle{\frac{p\cdot q}{p\cdot k} =\frac{Q^2}{2p\cdot k}}$ 
and $E$ is the energy of the incident neutrino (antineutrino) beam
in the laboratory system.

In realistic calculations the above cross section has then to be averaged
over the experimentally available neutrino spectrum, which is often
known with relatively poor precision.
In order to minimise the uncertainties induced by this averaging
procedure and, possibly, to enhance
the sensitivity to the axial strange contributions,
it is useful to consider appropriate ratios of cross sections,
which have the additional advantage of reducing 
nuclear structure effects,
when the target nucleons are bound into nuclei.

Two types of ratios are typically considered:
the NC over CC ratio, for both $\nu$ and $\overline{\nu}$ processes,
\begin{equation}
R_{NC/CC}(Q^2) = 
\frac{
\left(d\sigma/dQ^2\right)_{\nu(\overline{\nu})}^{NC}
}
{
\left(d\sigma/dQ^2\right)_{\nu(\overline{\nu})}^{CC}
}
\label{r_nc_cc}
\end{equation}
and the so--called NC proton over neutron ratio
\begin{equation}
R_{p/n}^{\nu}(Q^2) = 
\frac{
\left(d\sigma/dQ^2\right)_{\nu p}^{NC}
}
{
\left(d\sigma/dQ^2\right)_{\nu n }^{NC}
}\;.
\label{r_pn}
\end{equation}
The former has been measured (for total cross sections) in the
BNL--734 experiment at Brookhaven \cite{ahrens} and is currently
being considered for possible future measurements with the NuMi
beam at Fermilab \cite{morfinKEK}; 
for the latter, a proposal exists for an experiment to be done
at Los Alamos \cite{garLAMPF}, by measuring the ratio
of proton and neutron yields for neutrino quasi--elastic
scattering on carbon; however some preliminary results  
\cite{tay02} seem to indicate that the error bars are too
large to provide
a precise measurement of $G_A^s$. 

Another very interesting quantity is the NC/CC neutrino--antineutrino
asymmetry \cite{alb02,alb96,alb97}:
\begin{equation}
\Acal(Q^2)= \frac{\displaystyle{\left(\frac{d\sigma}{dQ^2}\right)^{NC}_{\nu} -
\left(\frac{d\sigma}{dQ^2}\right)^{NC}_{\nubar}}}
{\displaystyle{\left(\frac{d\sigma}{dQ^2}\right)^{CC}_{\nu} -
\left(\frac{d\sigma}{dQ^2}\right)^{CC}_{\nubar}}}\,,
\label{nuasymm}
\end{equation}
which, in terms of the single nucleon form factors, reads:
\begin{eqnarray}
\Acal_{p(n)}&=&\frac{1}{4 |V_{ud}|^2}
\left(\pm 1-\frac{\GAs}{\GACC}\right)
\left(\pm 1-2\sin^2\theta_W\frac{{\GMem}^{p(n)}}{\GMem^3} 
-\half\frac{\GMs}{\GMem^3}\right)
\nonumber\\
&=& 
\Acal_{p(n)}^0 \mp\frac{1}{8 |V_{ud}|^2}\frac{\GMs}{\GMem^3} 
\mp \frac{\GAs}{\GACC}\Acal_{p(n)}^0 \;.
\label{nuasymm2}
\end{eqnarray}
In the last expression only the linear terms in 
the strange form factors have been taken into account and
\begin{equation}
\Acal_{p(n)}^0= \frac{1}{4 |V_{ud}|^2}\left(1\mp 
2\sin^2\theta_W\frac{{\GMem}^{p(n)}}{\GMem^3} \right)
\label{nuasymm0}
\end{equation}
is the expected asymmetry when all the strange form factors 
are equal to zero.

\begin{figure}[t]
\centerline{\epsfxsize=3.5in\epsfbox{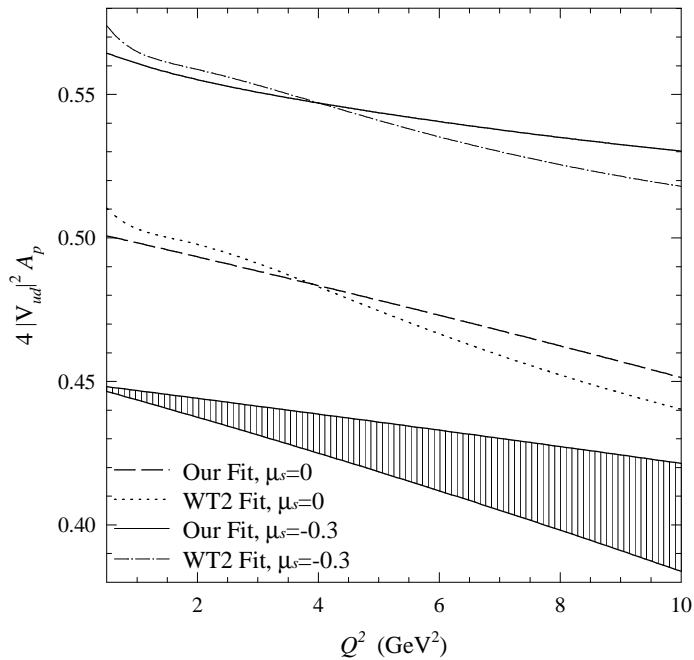}}   
\vskip 0.2 cm
\caption{
Plot of
$4|V_{ud}|^2{\Acal}_p$ as a function of $Q^2$. 
The shaded band represents
the uncertainty in the ``reference value'' $\Acal_p^0$ due to
the experimental errors on the magnetic form factors $G_M^{p,n}$.
The other four curves are calculated with $G_A^s(0) = g_A^s = -0.15$, 
while $G_M^s(0)= \mu_s=0$ for the
dotted and dashed curves, and $\mu_s = -0.3$ for the solid and dot
dashed ones. Two parameterisations of the electromagnetic form factors
are used, as explained in the text.}
\label{fig:1}
\end{figure}

An example of the dependence and sensitivity of the asymmetry (\ref{nuasymm}) 
on the different strange form factors is shown in fig. \ref{fig:1}, where
the uncertainty due to electromagnetic form factors has been illustrated
by considering two possible parameterisations, labelled here as
``our fit'' \cite{alb96} and WT2 \cite{WT}. Although it is not
extremely sensitive to the value of $g_A^s$, the interest of this
quantity stems from the fact that, as shown in eq. (\ref{nuasymm2}),
any deviation from the known reference value $\Acal_{p(n)}^0$
would be a proof of a non--negligible contribution of the strange
form factors, independently on the assumptions made for their
$Q^2$ dependence. In particular, if $G_M^s$ is known
with sufficient accuracy from P--odd electron scattering, it
would be possible to extract $G_A^s$ in a model independent way. 

\section{The BNL--734 experiment}
\begin{figure}[t]
\centerline{\epsfxsize=3.5in\epsfbox{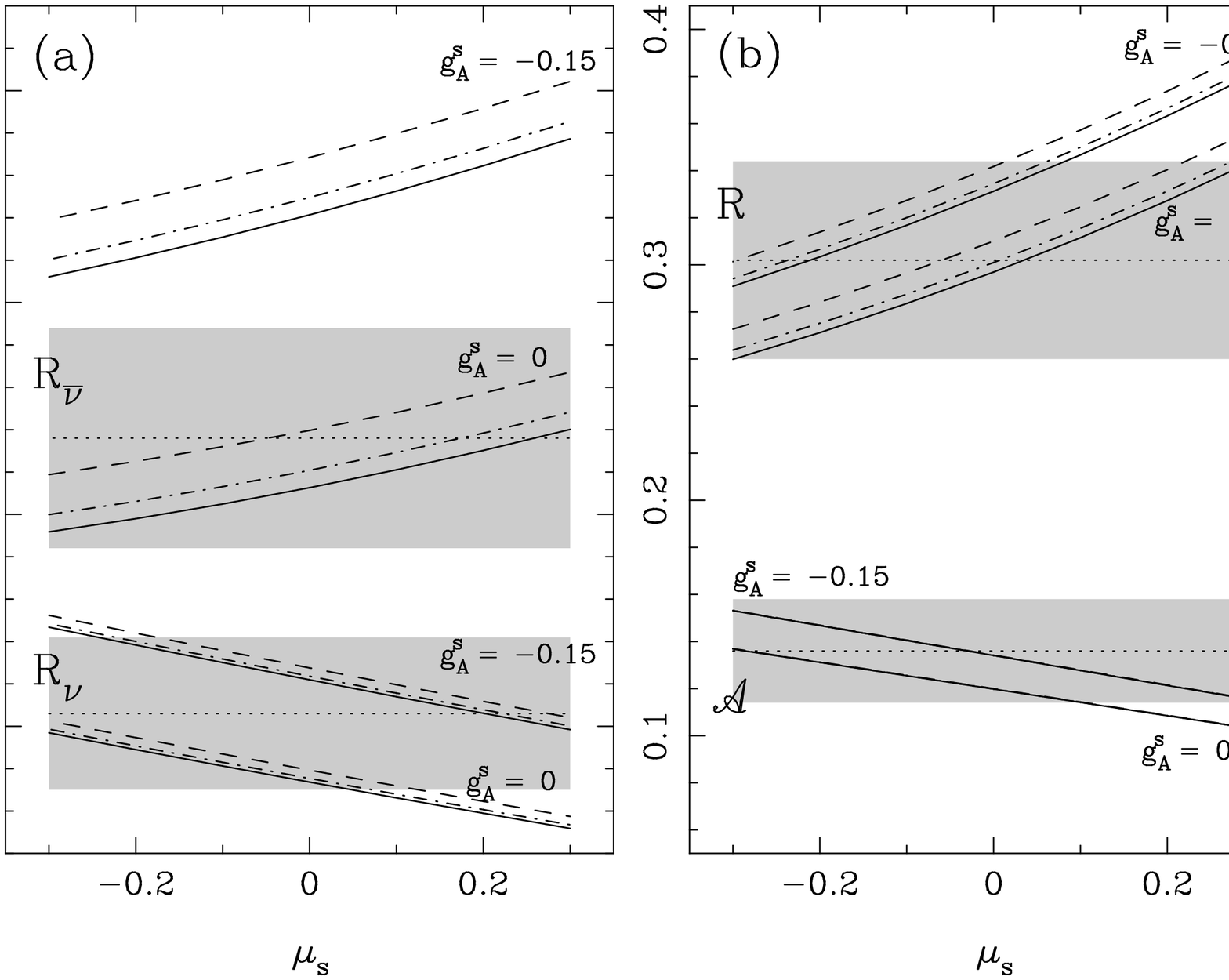}}   
\vskip -1 cm
\caption{The ratios (\ref{Rnu}), (\ref{Rnubar}), (\ref{RBNL})
and the folded asymmetry (\ref{eq.foldasymm})
as functions of $G_M^s(0)$, for different values of
$G_A^s(0)$, as indicated. 
The set of curves shown for each of these values correspond to
different choices of the strange electric form factor $G_E^s$.}
\label{fig:0}
\end{figure}

Up to now the most detailed study of strangeness effects in neutrino--nucleon
scattering has been done in the Brookhaven E--734 experiment
in 1987~\cite{ahrens}, using a wide band neutrino beam, 
with average energies
$1.3$ and $1.2$ GeV for neutrinos and antineutrinos, respectively, and a
170 ton high resolution liquid scintillator target--detector.
About 79\% of the target protons were bound in carbon and aluminum nuclei
and 21\% were free protons; Fermi motion and other nuclear effects were 
taken into account in the analysis of the data, in order to provide 
``equivalent free'' scattering data.
From an analysis of the measured, flux averaged, $\nu p$ 
and $\overline{\nu}p$ differential cross sections 
\begin{equation}
\langle
\frac{ d\sigma}{d Q^2}
\rangle^{NC}_{\nu (\overline{\nu})} =
\frac{ 
\displaystyle{
\int d 
E_{\nu (\overline{\nu})} 
\left(d\sigma/dQ^2\right)^{NC}_{\nu (\overline{\nu})} 
\Phi_{\nu(\overline{\nu})} 
\left( 
E_{\nu (\overline{\nu})} 
\right)
 }
}
{ 
\displaystyle{
\int d E_{\nu(\overline{\nu})}
\Phi_{\nu(\overline{\nu})} 
\left( 
E_{\nu (\overline{\nu})} 
\right)
} }\;,
\label{eq:BNLcross}
\end{equation}
an indication for a non zero isoscalar contribution to the axial
form factor was found, with $-0.25 \le G_A^s(0) \le 0$ at 90\% CL.
However, as confirmed by a subsequent and more precise re-analysis
\cite{garBNL}, a strong correlation between $G_A^s(0)$ and the dipole
cutoff mass $M_A$, employed to describe the $Q^2$--dependence of
both $G_A$ and $G_A^s$, was observed, 
which prevents an unambiguous extraction the axial 
strange form factor form the BNL data.

The BNL--734 experiment also measured the following ratios
of $Q^2$--integrated cross sections:
\bea
R_{\nu}&=& 
\frac{\langle\sigma(\nu_{\mu}p\to\nu_{\mu}p)\rangle}
{\langle\sigma(\nu_{\mu}n\to\mu^- p)\rangle}=0.153\pm 0.007\, (\mathrm{stat})\,
\pm 0.017\,(\mathrm{syst})
\label{Rnu}\\
R_{\nubar}&=& \frac{\langle\sigma(\nubar_{\mu}p\to\nubar_{\mu}p)
\rangle}
{\langle\sigma(\nubar_{\mu}p\to\mu^+ n)\rangle}=0.218\pm 0.012\,
 (\mathrm{stat})\,
\pm 0.023\,(\mathrm{syst})
\label{Rnubar}
\\
R &=& \frac{\langle\sigma(\nubar_{\mu}p\to\nubar_{\mu}p)\rangle}
{\langle\sigma(\nu_{\mu}p\to\nu_{\mu}p)
\rangle} =0.302\pm 0.019\, (\mathrm{stat})\,
\pm 0.037\,(\mathrm{syst})\,.
\label{RBNL}
\eea
which were later re-analysed in detail~\cite{albBNL}, in connection 
with the  flux averaged ``integrated'' neutrino--antineutrino asymmetry
derived from the above ratios through the relation
\begin{eqnarray}
\langle {\Acal^I} \rangle &=& 
\frac{
\langle \sigma 
\rangle_{\nu}^{NC} - 
\langle \sigma 
\rangle_{\overline{\nu}}^{NC}
}{
\langle \sigma 
\rangle_{\nu}^{CC} - 
\langle \sigma 
\rangle_{\overline{\nu}}^{CC}
}
\label{eq.foldasymm}
\\
&=& 
\frac{
\displaystyle{
R_{\nu} 
\left(1 - R\right)
 }}
{\displaystyle{
1 - R R_{\nu} /R_{\overline{\nu}}
}}
= 0.136 \pm 0.008 \,(\mathrm{stat})\,
\pm 0.019\,(\mathrm{syst})\,.
\nonumber
\end{eqnarray}
As illustrated in fig. \ref{fig:0}, 
the combined analysis of the above four quantities seems to
exclude large negative values of $G_A^s$ and to favour a 
negative value of $G_M^s$, but, in agreement with the previous findings
\cite{garBNL},
the error bars were found to be too large to allow any
definite conclusion.
The sensitivity of these observables to several other effects 
was also studied: while nuclear structure effects and uncertainties
in the neutrino flux were found to be largely reduced in all of them,
the sensitivity of the ratios (\ref{Rnu})--(\ref{RBNL}) to the
axial cutoff mass was shown to be still rather relevant.
On the contrary, the asymmetry (\ref{eq.foldasymm}) is practically
independent of $M_A$.

\section{Future perspectives at Fermilab}

The recent experimental results on neutrino oscillations
have motivated the realization
of new high intensity neutrino beams \cite{nu}, whose energy
spectra should be known with accuracies of a few percent.
Besides their primary goal to study neutrino physics, these future
experiments could be used to obtain new and much more accurate 
measurements of neutrino cross sections.
In particular the NuMi facility is currently under construction
at Fermilab \cite{NuMi}.
A high intensity, neutrino beam will be available in the next few years, 
in three possible configurations:
a low-energy beam, peaked at 3 GeV and with average energy
of about 6 GeV, a medium energy beam with peak and average energies
of about 6 and 7 GeV, respectively, and a high energy beam
with peak and average energies of about 7 and 11 GeV~\cite{morfin}.
The low energy configuration, in particular, looks very promising 
for accurate  studies of neutrino scattering processes. 
We have explored this possibility by using the expected flux to calculate 
the ratios (\ref{r_nc_cc}) and  (\ref{r_pn}) of flux--averaged neutrino
cross sections, studying their sensitivity to strange form
factors, as well as the effects of other possible 
theoretical and experimental uncertainties.
\begin{figure}[t]
\centerline{\epsfxsize=3.5in\epsfbox{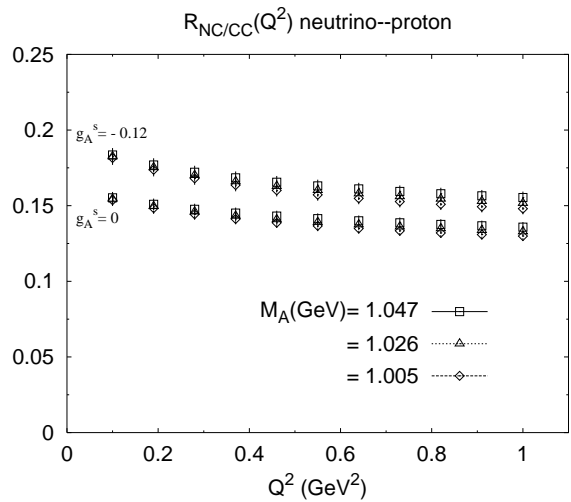}}  
\caption{Plot of the ratio $R_{NC/CC}(Q^2)$, obtained with 
the neutrino cross sections averaged over the low energy NuMi spectrum,
for different choices of $M_A$ and $g_A^s$.}
\label{fig:2}
\end{figure}

The NC/CC ratio for neutrino processes
is shown in fig.~\ref{fig:2}, for different choices
of the axial cutoff mass $M_A$ and of the strange axial 
constant $g_A^s$, as indicated. We have assumed that
this ratio could be measured with a 5\% accuracy, represented by the
small ``error band'' plotted for each calculated point.
We can see that, for the moderate $Q^2$ values represented here, 
the sensitivity of this ratio to $G_A^s$ is large enough
to allow a precise determination of it. 

\begin{figure}[t]
\centerline{\epsfxsize=3.5in\epsfbox{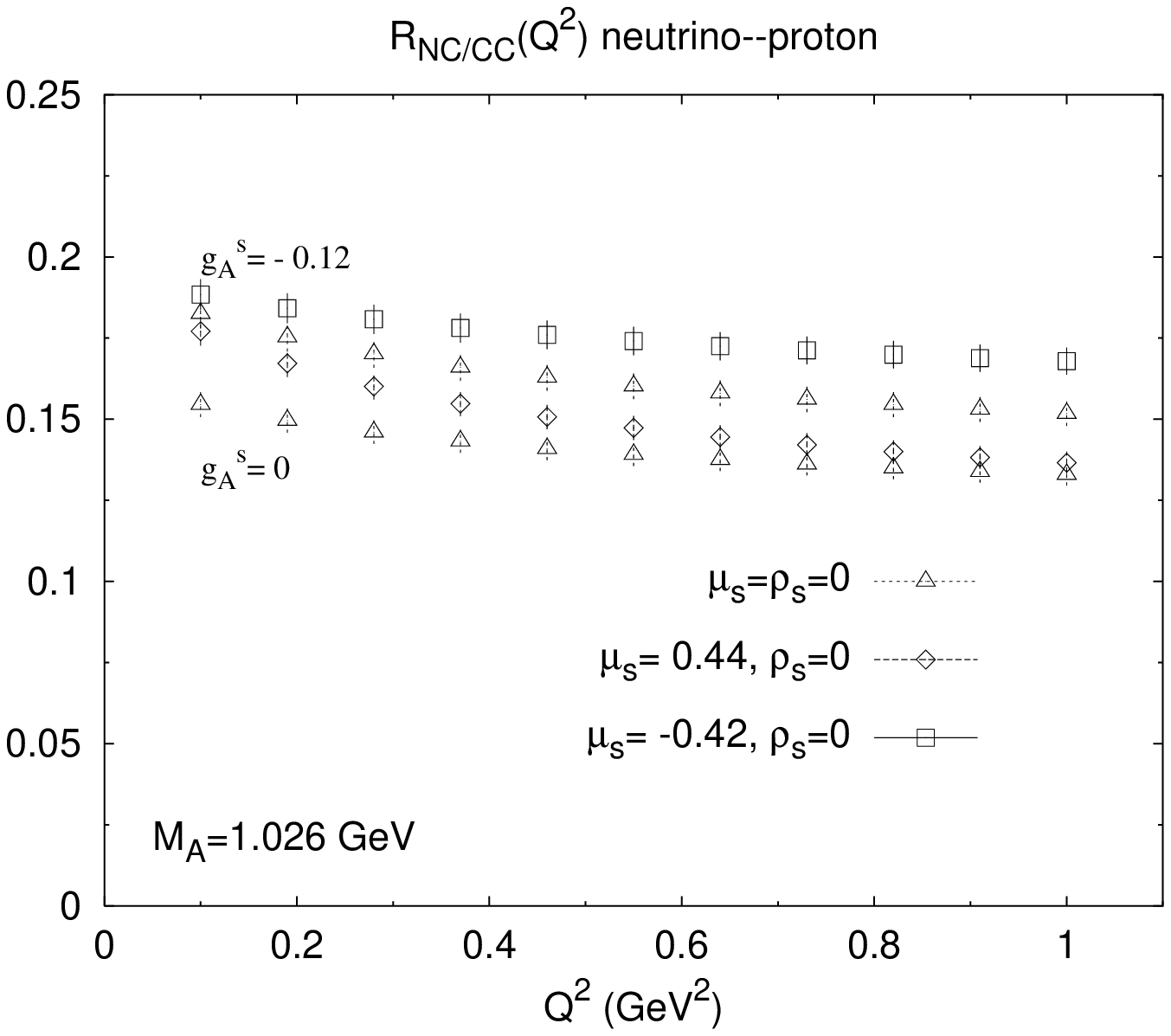}}   
\caption{Sensitivity of the neutrino ratio $R_{NC/CC}(Q^2)$ to the strange
magnetic form factor.}
\label{fig:3}
\end{figure}
Even if the effects of the strange vector currents in neutrino
scattering are usually smaller those of $G_A^s$, they can anyway
interfere with the extraction of the latter from the data.
We have thus studied the effects of the magnetic form factor $G_M^s$
on the NC/CC ratio, letting $G_M^s(0)=\mu_s$ vary in the rather
large range, corresponding to the error bar in 
the measurement of the SAMPLE collaboration~\cite{SAMPLE}. 
From the results shown in fig.~\ref{fig:3} it is seen that
a large positive value of $G_M^s$ would indeed almost compensate 
the effects of axial strangeness. This indicates that more precise
information on the strange magnetic form factor is 
needed as an input, in order to study $G_A^s$ with neutrino probes.
The effects on the ratio $R_{NC/CC}(Q^2)$ of the electric strange form
factor $G_E^s$ and those of the electromagnetic form factors have been
found to be very small and are not shown here.
Other possible sources of uncertainty in the extraction of $G_A^s$
have been also investigated:
nuclear effects were studied, by calculating the same 
type of ratio for quasi--elastic scattering on carbon nuclei, described
within the relativistic Fermi gas, while the sensitivity to the 
flux--averaging procedure was tested, 
by comparing the ratio of ``folded'' cross
sections with the same ratio at fixed $E_{\nu}$, for a few choices
of energy values. In both cases  no significant effects were found.
\begin{figure}[t]
\centerline{\epsfxsize=3.5in\epsfbox{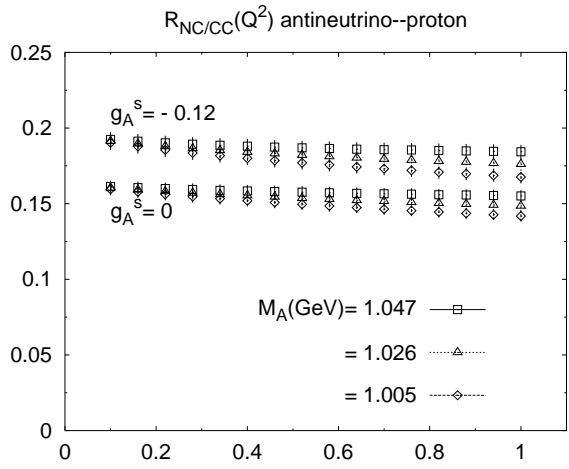}}   
\caption{Plot of the ratio $R_{NC/CC}$ for an hypothetical antineutrino
beam corresponding to 1\% of the NuMi $\nu$ spectrum.}
\label{fig:4}
\end{figure}

Even if antineutrinos will not be immediately available in the low energy 
NuMi beam, it is interesting to consider the sensitivity to strangeness of the 
corresponding NC/CC ratio, which we have calculated by assuming a 
$\overline{\nu}$ beam corresponding to 1\% of the NuMi neutrino spectrum.
The results of this calculation are shown in fig.~\ref{fig:4}, where
the effects of $g_A^s$ appear to be similar to those for the neutrino case,
with a slightly larger sensitivity to $M_A$.
Results similar to the case of neutrinos are also obtained for
the sensitivity to the vector strange form factors, although in the case of 
$\overline{\nu}$ larger effects of the $G_E^s$ are observed, stressing again
the importance of obtaining precise complementary results from electron 
scattering.
\begin{figure}[t]
\centerline{\epsfxsize=3.5in\epsfbox{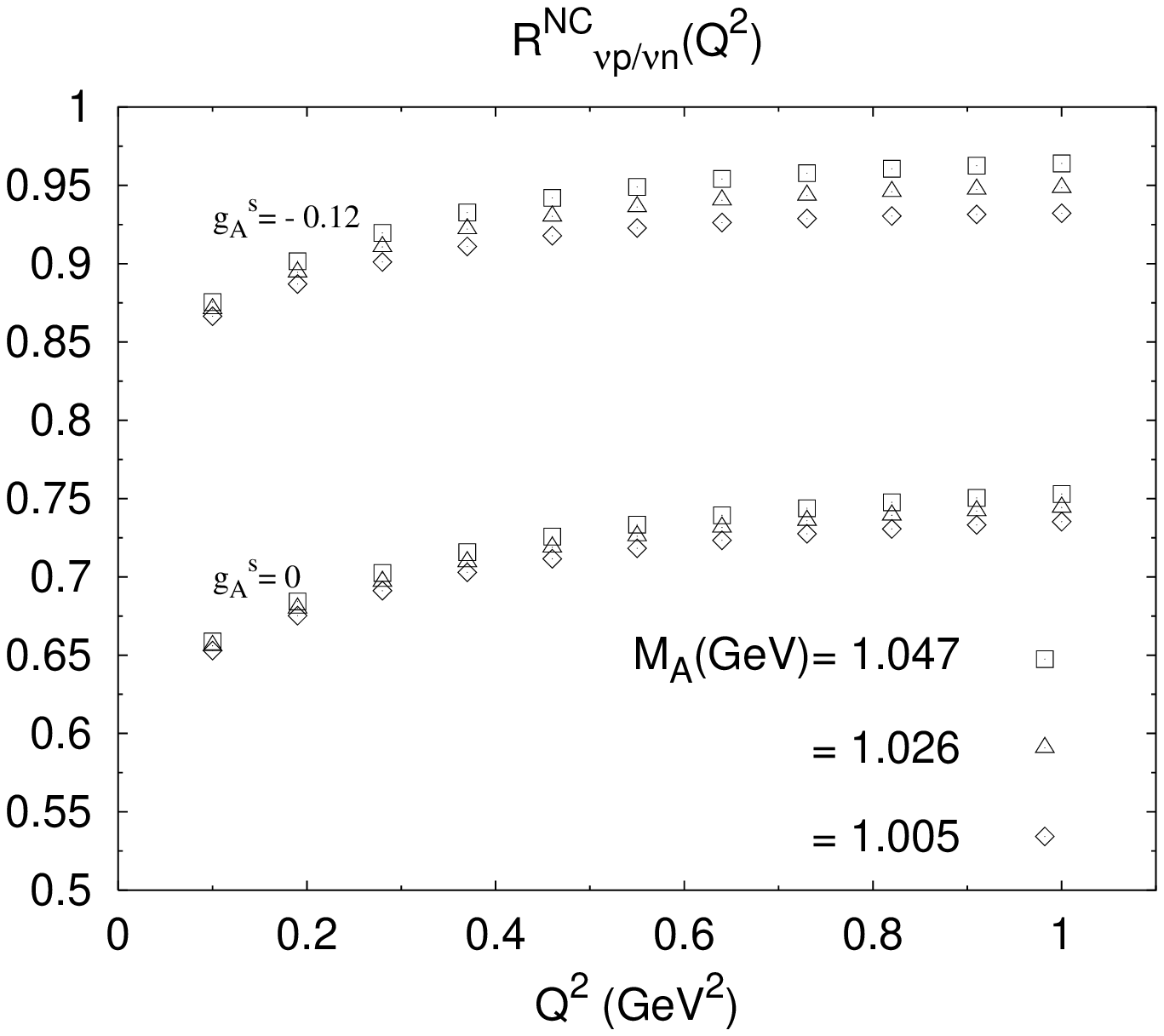}}   
\caption{Plot of the ratio $R_{p/n}^{\nu}$ for the scattering of the NuMi
low energy neutrinos on $^{12}C$}
\label{fig:5}
\end{figure}

\begin{figure}[t]
\centerline{\epsfxsize=3.5in\epsfbox{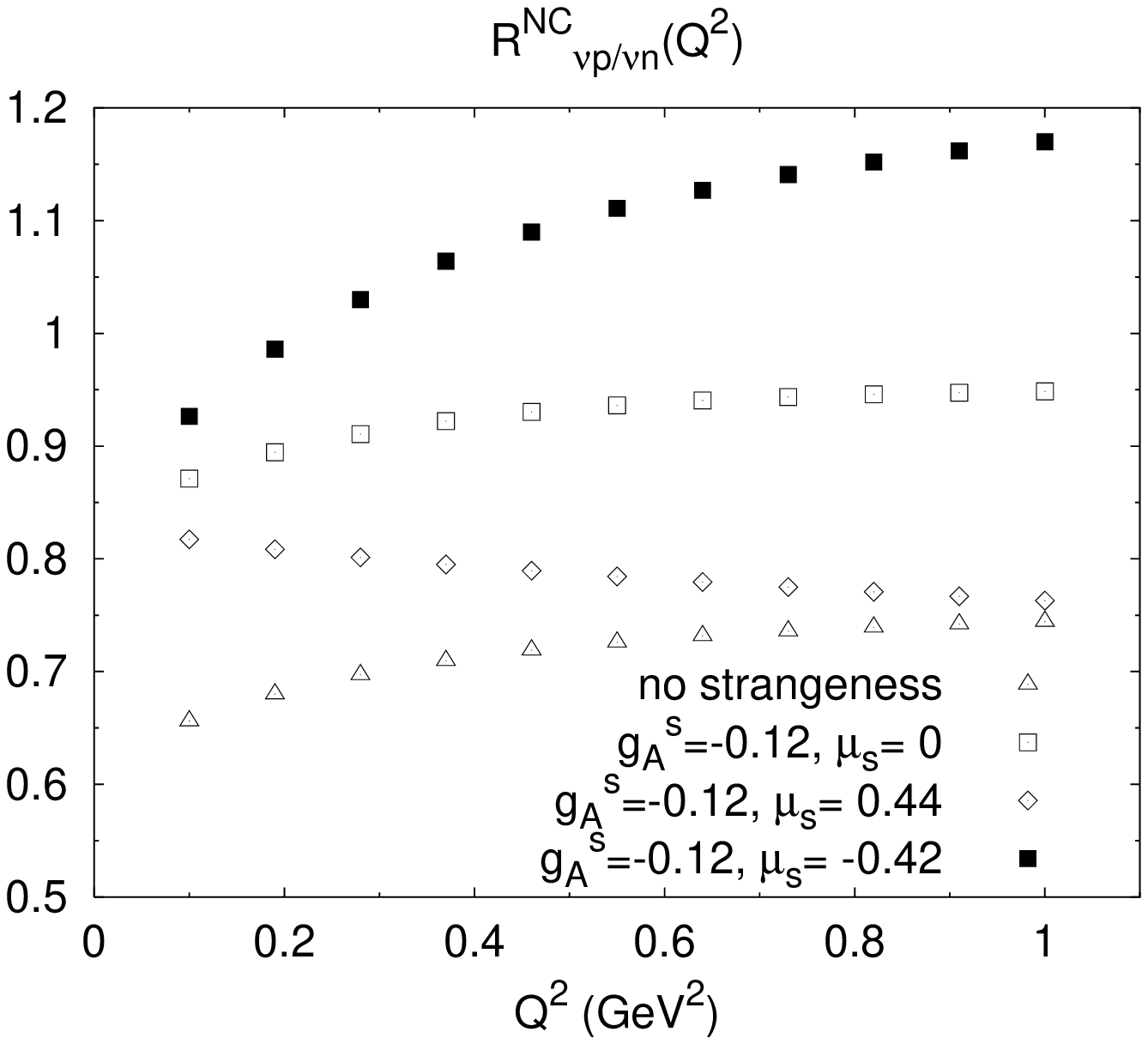}}   
\caption{Same as in fig. \ref{fig:5}, but for different values
of $\mu_s=G_M^s(0)$.}
\label{fig:6}
\end{figure}

Finally, we have considered the proton over neutron ratio (\ref{r_pn})
of flux averaged neutrino cross sections, in the case of quasi--elastic 
scattering on $^{12}C$, described within the Fermi gas model (again,
no difference is obtained with respect to the corresponding ratio of 
free nucleon cross sections)
\footnote{In this case $Q^2$ is to be interpreted as an ``equivalent
free momentum transfer'', $Q^2 \equiv 2 M T_N$, $T_N$ being the outgoing
nucleon kinetic energy.}. 
The sensitivity of this ratio to $G_A^s$ and $M_A$ as well as to $G_M^s$
is shown in figs.~\ref{fig:5} and \ref{fig:6}, respectively.
We can see that the effects of the axial strange form factor are very large
and are not significantly masked by the uncertainty on $M_A$, while,
again, the present large error band on $G_M^s$ would not allow
a precise determination of $G_A^s$ from this measurement.

Similar conclusions for all the above considered observables
can be drawn when considering the corresponding ratios 
of $Q^2$--integrated cross sections.

\section{Elastic Scattering on $S=T=0$ nuclei}

Before concluding we want to briefly discuss the possibility of using
neutrino scattering to study nuclear strange form factors,
by measuring the cross sections for the elastic scattering of
(anti)neutrinos on $S=T=0$ nuclei,
\begin{equation}
\nu~(\nubar) + A \longrightarrow \nu~(\nubar) + A \;. 
\label{nuAscat}
\end{equation}
In this case, both
the axial current $A_{\alpha}$ and the isovector part of the polar vector
weak neutral current, $V_{\alpha}^3 (1-2\sin^2(\theta_W))$,
do not contribute to the cross sections, whose expressions are given by:
\begin{equation}
\frac{d\sigma_{\nu}}{dQ^2}=\frac{d\sigma_{\nubar}}{dQ^2}=
\frac{G_F^2}{2\pi}\left(1-\frac{p\cdot q}{M_A E} -\frac{Q^2}{4E^2}\right)
\left[F^{NC}(Q^2)\right]^2\;.
\label{nucross00}
\end{equation}
Here $E$ is the neutrino energy in the laboratory system and 
$F^{NC}(Q^2)$ is the 
NC nuclear elastic form factor, which can be expressed in terms of 
the isoscalar nuclear electromagnetic
form factor $F(Q^2)$ and of an unknown  strange contribution 
$F^s(Q^2)$, in the following form:
\begin{equation}
F^{NC}(Q^2)= -2\sin^2\theta_W F(Q^2) -\half F^s(Q^2) \;.
\end{equation}

The measurement of the elastic cross section (\ref{nucross00}) requires 
the detection of the small recoil energy of the final nucleus, and it is 
thus a very challenging experimental task. However it could provide
very important information on the nuclear strange form factor.
In fact,  if the electromagnetic form factor $F(Q^2)$ is obtained from
a measurement of the cross section for the elastic scattering of 
unpolarised electrons,
\begin{equation}
\frac{d\sigma_e}{dQ^2}=\frac{4\pi\alpha^2}{Q^4}
\left(1-\frac{p\cdot q}{M_A E} -\frac{Q^2}{4E^2}\right)
\left[F(Q^2)\right]^2\,,
\label{eAelas}
\end{equation}
then $F^s(Q^2)$
could be obtained directly from measured cross sections:
\begin{equation}
F^s(Q^2)= \pm 2 F(Q^2)\left\{\left(\frac{2\sqrt{2}\pi\alpha}{G_F Q^2}\right)
\sqrt{\frac{(d\sigma_{\nu}/dQ^2)}{(d\sigma_e/dQ^2)}} \mp 2\sin^2\theta_W
\right\}\;,
\label{sffnucl2}
\end{equation}
and thus in a completely model independent way. 

\section{Conclusions}
In conclusion we believe that neutrino scattering is a very important
tool to study strangeness contributions to the structure of the
nucleon. New high intensity neutrino beams, available in the near future,
and improvements in the precision of the measurements of the vector strange
form factors in polarised electron scattering, could allow
a definite determination of the axial strange form factor $G_A^s$.
We would like to stress, that, although seldom available, antineutrino
beams, would offer the possibility of accessing additional complementary
information on the nucleon axial strangeness; in particular 
the measurement of the NC/CC neutrino--antineutrino asymmetry would
provide a model independent method to determine $G_A^s$.



\end{document}